\documentclass[proceedings]{JHEP} % 10pt is ignored!

\newcommand{\tr}[1]{\,{\rm tr}\,#1\,}

\def\be{\begin{equation}}

\def\ee{\end{equation}}
\def\bea{\begin{eqnarray}}
\def\eea{\end{eqnarray}}

\def\a{\alpha}

\def\n{\nabla}

\def\d{\delta}

\def\l{\left(}
\def\r{\right)}
\def\p{\partial}

\newcommand{\cN}{{\cal N }}
\newcommand{\cL}{{\cal L }}
\newcommand{\non}{\nonumber}
\conference{Quantum Aspects of Gauge Theories, Supersymmetry and
Unification }

\title{Scalar Quartic Effective Action on AdS$_5$}

\author{Gleb Arutyunov$^{a}$, Sergey Frolov$^{b}$ 
\\
$^a$Sektion Physik der
Ludwig-Maximilians-Universit\"at \\
Theresienstra{\ss}e 37, D-80333 M\"unchen, Germany
\\
E-mail: \email{ arut@theorie.physik.uni-muenchen.de } 
$~~~~~~~~~~~~~~$
\\
$^b$ Department of Physics and Astronomy, \\
University of Alabama, Box 870324, \\
Tuscaloosa, Alabama 35487-0324, USA 
\\
E-mail: \email{frolov@bama.ua.edu } 
}

\abstract{We review the recent results concerning 
the computation of cubic and quartic couplings of scalar fields 
in type IIB supergravity on $AdS_5\times S^5$ background that
are dual to (extended) chiral primary operators in $\cN=4$ SYM$_4$.
We discuss the vanishing of certain cubic and quartic couplings 
and non-renormalization property of corresponding correlators in
the conformal field theory.} 

\begin{document}

\section{Introduction}
The AdS/CFT duality \cite{M,GKP,W} provides the holographic relation 
between ${\cal N}=4$ supersymmetric 
Yang-Mills theory in four dimension (SYM$_4$)
and type IIB supergravity on $AdS_5\times S^5$ background.
An important class of operators in SYM$_4$ playing the role
of the BPS states is given by the chiral primary operators (CPO). 
Thus, the AdS/CFT correspondence
gives a remarkable opportunity to study the dynamics 
of CPOs (BPS states) at the strong coupling regime.

The power of the superconformal symmetry in four 
dimensions is not enough to supply the Operator Product Expansion (OPE)
of CPOs $~~~$ with the ring structure. In this respect the knowledge 
of correlation functions  of CPOs provides us with  new dynamical 
information about $\cN=4$ SYM$_4$ at the strong 't Hooft coupling 
(spectrum of operators in the OPE algebra of CPOs, their 
anomalous dimensions and so on). Comparison of the correlation
functions at strong and week couplings brings also insight 
previously unknown non-renormalization theorems.   

An important problem that has not been $~~~~$ solved 
yet is the computation 
of the 4-point correlation functions of CPOs in the supergravity 
approximation. To compute a 4-point correlation 
function in $\cN=4$ SYM$_4$ one has 
to derive the relevant part of the gravity 
action on $AdS_5$ background up to the fourth order and then 
to compute the on-shell value of this action 
(with Dirichlet boundary conditions on gravity fields). 
By perturbation theory the second step 
can be represented as evaluation of exchange and contact Feynman diagrams.
It is the absence of the well-defined 5d gravity action 
for fields that correspond to CPOs in SYM$_4$ that makes  
computation of the correlation functions of CPOs
a non-trivial problem. So far the only known 
examples here are the 4-point 
functions  of operators $\tr (F^2+\cdots )$ 
and $\tr (F\tilde{F} +\cdots )$ that 
on the gravity side correspond to massless modes 
of dilaton and axion fields \cite{LT1,HFMMR},
where the relevant part of the gravity action was known.  
These operators, however, are descendents of the CPOs 
that brings considerable complications both 
in perturbative analysis of the correlation functions, 
and in the study of their OPE from AdS gravity \cite{HMMR}.   

Recently we have constructed the quartic effective 5d 
action for scalar fields $s^I$
dual at linear order to CPOs \cite{AF6},
which allows one to compute  4-point 
functions of {\it any} (extended) CPOs 
in the supergravity approximation. Here we 
present a brief review of the corresponding 
results. The quartic couplings of the 5d action possess a number 
of remarkable properties. In particular they admit 
the consistent Kaluza-Klein 
reduction to the fields from the massless graviton multiplet
and they vanish in the so-called extremal and sub-extremal cases. 
Each of these properties implies the existence 
of the non-renormalization theorem
for the corresponding correlators 
of CPOs in SYM$_4$.

\section{CPOs, field redefinition and operator mixing} 
In $\cN=4$  SYM$_4$ there are ``short'' or 
chiral multiplets generated by primary operators:
\bea
O^I(x )=\frac{(2\pi )^k}{\sqrt{k\lambda^k}}C^I_{i_1\cdots i_k}
\tr (:\phi^{i_1}(x )\cdots \phi^{i_k}(x ):),
\non
%\la{cpo}
\eea
where $C^I_{i_1\cdots i_k}$ are totally symmetric traceless rank $k$ 
orthonormal tensors of $SO(6)$: 
$$\langle C^IC^J\rangle =C^I_{i_1\cdots i_k}C^J_{i_1\cdots i_k}=\delta^{IJ},$$ 
$\phi^{i}$ are scalars of SYM$_4$, the notation $:A_1\cdots A_n:$ 
is used to denote the 
normal-ordered product of the operators $A_i$ and $\lambda=g_{YM}^2N$
is the 't Hooft coupling.

Eight from sixteen supercharges annihilate $O^I(x )$ 
while the other eight ones 
generate chiral multiplets. A fundamental 
property of CPOs is that their conformal dimensions are protected. 
Thus, they may be regarded as BPS 
states preserving 1/2 of the 
supersymmetry.

The two- and three-point functions of CPOs computed in free theory (in the leading order in 
$1/N$) are \cite{LMRS}
\bea
\langle O^I(x )O^J(y )\rangle &=&\frac{\delta^{IJ}}{|x -y |^{2k}},
\non
%\la{cpo2}
\eea
\bea
\non
\langle O^{I_1}(x_1 )O^{I_2}(x_2 )O^{I_3}(x_3)\rangle 
=\frac 1N
\frac{\sqrt{k_1k_2k_3}C^{I_1I_2I_3}}
{x_{12}^{2\a_3}x_{23}^{2\a_1}x_{13}^{2\a_2}},
%\la{cpo3}
\eea
where $x_{ij}=x_i-x_j$,  
$\a_i =\frac 12 (k_j+k_l-k_i)$, $j\neq l\neq i$, and 
$C^{I_1I_2I_3}$ is the unique $SO(6)$ invariant obtained 
by contracting $\a_1$ indices between $C^{I_2}$ and  $C^{I_3}$, 
$\a_2$ indices between $C^{I_3}$ and  $C^{I_1}$, and
$\a_3$ indices between $C^{I_2}$ and  $C^{I_1}$.

Analysis of the superconformal transformations in SYM$_4$ 
allows one to conclude that CPOs $O^I(x )$ are dual to the scalars 
$s^I$ of IIB supergravity 
compactified on $AdS_5\times S^5$. Fields
$s^I$ appear in spectrum of the compactified theory 
as a linear combination  of the trace of the graviton 
on $S^5$ and of the 5-form field strength on $S^5$.
Index $I$ runs the KK tower of scalar spherical harmonics.

In fact there are only two ways to tackle the problem
of constructing an effective 5d action for the scalars $s^I$.
The first way is to compactify on $AdS_5\times S^5$
the covariant action for type IIB supergravity by \cite{DLS}.
Since the ten-dimensional theory contains a four-form with the self-dual 
field strength the manifest Lorenz covariance is achieved by
introducing an auxiliary scalar field $a$. Fixing $a$
to be some function of space-time variables and breaking 
thereby an additional $~~~$ gauge invariance associated with this field 
one gets a non-covariant action. The covariance w.r.t. the background
isometry group is then restored by means of introducing some additional 
non-propa\-gating fields. Namely this way was used 
in \cite{AF3} to derive the quadratic action for physical fields 
of IIB supergravity on  $AdS_5\times S^5$ background. 
However, at the level of cubic and quartic actions
the problem of solving the non-covari\-ant constraints imposed by 
gauge symmetries becomes extremely complicated. 

The second approach deals with the covariant equations of motion of 
IIB supergravity. The basic strategy here is to 
find quadratic and cubic corrections to 
equations of motion of gravity fields 
by decomposing the covariant equations 
up to the third order. The main problem here 
is that equations such obtained are 
non-Lagrangian and one has to perform
very complicated and fine analysis 
to reduce them to the Lagrangian form. 
In what follows we undertake the second approach.

The way to bring the equation of motion of IIB supergravity 
to the Lagrangian form is to perform non-linear redefinitions 
of the original fields $s^I$ in terms of the new fields ${s'}^I$.
Here we restrict our attention to fields $s^I$ but 
it should be noted that all other gravity fields require 
redefinitions of a similar type. Performing appropriate 
redefinitions one gets rid of the higher-derivative terms 
and simultaneously obtains the Lagrangian equations 
(see \cite{LMRS,AF6,AF5,Lee}).
Hence, in spite of the fact that the fields $s^I$ correspond 
to the simplest single-trace operators $O^I(x )$ in the Yang-Mills 
theory they are rather unnatural in the gravity description.
 
One may then wonder if the correspondence $O^I\leftrightarrow {s'}^I$
still holds. It turns out that in the extremal case, e.g., $k_3=k_1+k_2$,
3-point correlation functions of the corresponding CPOs computed 
in SYM$_4$ are non-singular when $x_1\to x_2$, while on the gravity side  
the cubic couplings for ${s'}^I$ vanish leading to the vanishing 
of the correlation functions. One natural conjecture \cite{AF3}
is that redefinitions of the gravity fields lead to the change
of the basis of CPOs in conformal field theory. We will widely 
refer to the new basis as to the {\it extended} CPOs. Thus, 
redefined fields ${s'}^I$  correspond to the extended CPOs.
It is possible to construct the extended CPOs explicitly.
Define the following operators in SYM$_4$:
\bea
\tilde {O}^{I_1}(x )&=&O^{I_1}(x ) 
\non  \\ 
&-&\frac {1}{2N}\sum_{I_2+I_3=I_1}
{\bf C}^{I_1I_2I_3}:O^{I_2}(x )O^{I_3}(x ):, 
\non
\eea
where ${\bf C}^{I_1I_2I_3}=\sqrt{k_1k_2k_3}C^{I_1I_2I_3}$.
In the large $N$ limit these
operators have the normalized two-point functions,
the same 3-point functions as CPOs if none of the relations
\bea
k_1+k_2=k_3,\quad k_2+k_3=k_1,\quad k_3+k_1=k_2,
\non
\eea
is satisfied, and vanishing 3-point functions otherwise. 
Thus, the extended CPOs are constructed by mixing 
single- with double-trace operators. Still
the expression for $\tilde {O}$ is incomplete 
and it should be further extended by normal-ordered 
products of CPOs and their descendents. 

In principle it seems 
possible to find an action for the scalars dual to 
CPOs by performing the field
redefinitions reversed to the ones used to reduce the equations of
motion to a Lagrangian form. The reversed transformations should be
made at the level of the quartic action, but not at the level of equations
of motion. However, the resulting action for 
the new scalars will be much more complicated and will contain 
higher-derivative terms with six derivatives. 
It's worth noting that the equations of motion derived from the new action
certainly differ from the original ones despite the fact
that one made reversed transformations. 
Hopefully the coincidence of the correlation functions 
of CPOs and of extended CPOs for generic values of conformal dimensions 
and the existence of the analytic continuation procedure advocated 
in \cite{LT2},\cite{DFMMR} allows one to find the correlation
functions of any CPOs by working with the gravity action 
for redefined fields.  
 
\section{Effective 5d action}
The result of our study in \cite{AF6} is the effective 5d action 
for scalars $s^I$
needed to compute 4-point correlation functions of {\it any}
(extended) CPOs in $\cN=4$ SYM$_4$ at large 't Hooft coupling:
\bea
\non
&&S(s)=\frac{4N^2}{(2\pi )^5}\int d^{5}x\sqrt{-g_a} ~\times \\ 
&& \biggl (\sum_i (\cL_2(\Phi_i)+\cL_3(\Phi_i))
+\cL_4^{(0)}+\cL_4^{(2)}+\cL_4^{(4)}\biggr ),
\non
%\la{action}
\eea
where $g_a$ stands for the determinant of the AdS-metric. 
Here $\Phi_i$ denotes one of the fields from the following set
$$
(s^I,t^I,\phi^I,\phi_{ab}^I,A_a^I,C_a^I).
$$
In fact this are all the fields that appear in cubic interaction vertices
$\cL_3(\Phi_i)$ containing two fields $s^I$.  
Here the quadratic terms are given by \cite{AF3}
\bea
&&\cL_2(s)=c_s\left( -\frac12 \n_as_k\n^as_k
-\frac12 m^2s_k^2\right),
\non \\
&&\cL_2(t)=c_t
\left( -\frac12 \n_at_k\n^at_k
-\frac12 m_t^2t_k^2\right), 
\non \\
&&\cL_2(\phi )= -\frac14 \n_a\phi_k\n^a\phi_k -\frac14 f(k)\phi_k^2 ,
\non \\
&&\cL_2(\varphi_{ab})=
-\frac{1}{4}\n_c\varphi_{ab}^k\n^c\varphi^{ab}_k
+\frac{1}{2}\n_a\varphi^{ab}_k\n^c\varphi_{cb}^k \non \\
&&-\frac{1}{2}\n_a\varphi_c^{ck}\n_b\varphi^{ba}_k  
+\frac{1}{4}\n_c\varphi_a^{ak}\n^c\varphi^{bk}_b \non \\
&&+\frac{1}{4}(2-f(k))\varphi_{ab}^k\varphi^{ab}_k+
\frac{1}{4}(2+f(k))(\varphi_{a}^{ak})^2 ,
\non \\
&&\cL_2(A_a)=c_A  
\left(-\frac14 (F_{ab}(A^k))^2
-\frac12 m_A^2(A_a^k)^2\right) ,
\non \\
&&\cL_2(C_a)=c_C \left(-\frac14 (F_{ab}(C^k))^2
-\frac12 m_C^2(C_a^k)^2\right) ,
\non 
\eea
where the masses of the particles are 
\bea
&&m^2= k(k-4),\quad m_t^2=(k+4)(k+8), \non\\
&&m_A^2=k^2-1, \quad m_C^2= (k+3)(k+5),\non\\
&&m_\phi^2=m_\varphi^2=f(k)=k(k+4), \non
\eea
$F_{ab}(A)=\p_aA_b-\p_bA_a$,
and the normalization constants depending on $k$ are found to be 
\bea
&&c_s=\frac{32k(k-1)(k+2)}{k+1},~~c_A=\frac{k+1}{2(k+2)}, \non \\
&&c_t=\frac{32(k+2)(k+4)(k+5)}{k+3},~~ c_C=\frac{k+3}{2(k+2)}. \non
\eea
To simplify the notation we denote here $s^{I_1}$ as $s_k$
or simply as $s^1$. The index $I\equiv I(k)$ runs the basis of  
a representation of $SO(6)$ specified by $k$. 
Terms $\cL_2$ above represent the standard quadratic 
Lagrangians for fields of different spins.
  
The cubic terms were found in \cite{LMRS,AF5,Lee}, and may be written 
as follows
\bea
&&\cL_3(s)= S_{123}~s^{1}s^{2}s^{3},
\non\\
\non
&&\cL_3(t)=T_{123}~s^{1}s^{2}t^{3},\\
&&\cL_3(\phi )=\Phi_{123}~s^{1}s^{2}\phi^{3},\non\\
\non
&&\cL_3(\varphi_{ab})=G_{123}
\l \varphi_{ab}^{3}T^{ab~12}
+\frac{f_3}{4}s^{1}s^{2}\varphi_c^{c 3}\r \\
&&\cL_3(A_a)= A_{123}~
s^{1}\n^a s^{2}A_a^{3},\nonumber\\
&&\cL_3(C_a)= C_{123}~
s^{1}\n^a s^{2}C_a^{3},\nonumber
\eea
Here to describe the interaction 
with the massive graviton $\varphi_{ab}^I$ we introduced 
the notation
\bea
\non
T_{ab}^{12}&=&\n^a s^{1}\n^b s^{2} \\
\non
&-&\frac{\d^{ab}}{2} \left( \n^c s^{1}\n_c s^{2}+
\frac12 m_{12}^2s^{1}s^{2}\right),
\eea
where the ``mass'' matrix is $m_{12}^2=m^2_1+m^2_2$.
Here the summation over $I_1,I_2,I_3$ is assumed and $f_3\equiv f(k_3)$.
For the explicit values of the cubic couplings see \cite{AF5}.

Finally the quartic terms recently found in \cite{AF6} are given by 
\bea
&&\cL_4^{(0)}=S_{1234}^{(0)}~s^{1}s^{2}s^{3}s^{4};
\non\\
&&\cL_4^{(2)}=
\l S_{1234}^{(2)}
+A_{1234}^{(2)}\r s^{1}\n_as^{2}s^{3}\n^as^{4},
\non
\eea
with the symmetry
\bea
S_{1234}^{(2)}&=&S_{2134}^{(2)}=S_{3412}^{(2)},
\non\\
A_{1234}^{(2)}&=&-A_{2134}^{(2)}=A_{3412}^{(2)};
\non
\eea
and by 
\bea
\cL_4^{(4)}&=&
\l S_{1234}^{(4)}
+A_{1234}^{(4)}\r \non \\
&\times& s^{1}\n_as^{2}\n_b^2(s^{3}\n^as^{4}),
\non
\eea
where
\bea
S_{1234}^{(4)}&=&S_{2134}^{(4)}=S_{3412}^{(4)},
\non\\
A_{1234}^{(4)}&=&-A_{2134}^{(4)}=A_{3412}^{(4)}.
\non
\eea
Explicit values of the quartic couplings are given in \cite{AF6}.
Equations of motion that follow from this action are related 
with original non-Lagrangian equations of IIB supergravity 
on the $AdS_5\times S^5$ background by a chain of 
{\it field redefinitions} and by the usage of {\it hidden relations}
between different kinds of spherical harmonics.
One of the novel features of the found couplings is the 
presence of the 2- and 4-derivative terms that can not be removed 
by any field redefinition without spoiling the minimal character of the 
interaction at the cubic level.  

\section{Reduction to the gauged ${\cal N}=8$ 5-dimensional supergravity} 
The quartic couplings we found allow us to study the problem of
the consistency of the Kaluza-Klein (KK) reduction down to five dimensions.
It is customarily believed that the $S^5$ compactification of type IIB 
supergravity admits a consistent truncation to the massless multiplet, 
which can be identified with the field 
content of the gauged ${\cal N}=8$, $d=5$ supergravity \cite{PPN,GRW}.
Consistency means that there is no term linear in 
massive KK modes in the untruncated supergravity action, 
so that all massive KK fields can be put to zero without any
contradiction with equations of motion.
From the AdS/CFT correspondence point of view the consistent 
truncation implies that $any$ $n$-point correlation function 
of $n-1$ operators dual to the fields from the massless multiplet
and one operator dual to a massive KK field vanishes because, as one can 
easily see there is no exchange Feynman diagram in this case.   

Considering explicit expressions for 
the cubic couplings found in \cite{AF5, Lee} one sees immediately  
that they obey the consistency condition allowing therefore  
truncation to the fields from the massless multiplet at the level 
of the cubic action. 

The truncation problem for the quartic couplings is a little bit more 
sophisticated.
Recall that the gauged ${\cal N}=8$ five-dimensional supergravity 
has in particular
42 scalars with 20 of them forming the singlet of the global 
invariance group $SL(2,{\bf R})$. 
These 20 scalars comprise the ${\bf 20}$ irrep. of $SO(6)$ and 
correspond to the IIB supergravity fields $s^I$ with $k=2$.
The five-dimensional scalar Lagrangian consists of the kinetic 
energy and the potential. 
The maximal number of derivatives appearing in the Lagrangian is 
two and that is due to 
the non-linear sigma model type kinetic energy. We have however found 
the quartic 4-derivative vertices that can not be shifted away by any 
field redefinition. Thus, a highly non-trivial check of the relation
between the compactification of the ten-dimensional theory and the gauged
supergravity in five dimensions as well of the results obtained    
consists in showing that the 4-derivative vertices vanish for the modes from 
the massless multiplet. It turns out that this is indeed the case. 
Moreover, after an additional simple 
field redefinition
the quartic vertices we found indeed vanish 
when one of the four fields is not from the massless multiplet,
proving thereby the consistency of the reduction at the level of the  
quartic scalar couplings. This in particular provides an additional
argument that the scalars $s^I$ (and, in general, any supergravity field) 
correspond not to CPOs but rather to extended CPOs.
Indeed, if we assume that the consistent truncation takes place at
all orders in gravity fields, we get that correlators of the form
$\langle O_2^{I_1}O_2^{I_2}\cdots O_2^{I_{n-1}}O_k^{I_n}\rangle$ 
vanish for $k\ge 3$.
This is certainly not the case for single-trace CPOs, and we are forced
to conclude once more that supergravity fields are in general dual
to extended operators which are admixtures of single-trace operators 
and multi-trace ones.\footnote{Note that the lowest modes $s_2$ may be dual
only to single-trace CPOs. It is possible that 
any field from the massless supergravity multiplet is dual to a single-trace
operator.} 
Since an extended operator is uniquely determined
by a single-trace one, it is natural to assume that if
a correlation function of extended operators vanishes then
there exists a kind of a non-renormalization theorem for an analogous 
correlation function of single-trace operators. 
If we further assume that type IIB string theory on $AdS_5\times S^5$ 
respects the consistent truncation, then the vanishing of $n$-point 
correlation functions of $n-1$ extended operators dual to
the supergravity modes from the massless multiplet, and one 
extended operator dual to a massive KK mode seems to imply that \cite{AF6}

{\it  at large $N$ the $n$-point functions of the 
corresponding single-trace 
operators are independent of 't Hooft coupling $\lambda =g_{YM}^2 N$.}

If the consistent truncation is valid at quantum level, that 
seems to be plausible because of a large amount of supersymmetry,
then these $n$-point functions are independent of $g_{YM}$ for any $N$.

In particular this conjecture is applied to 
$n$-point functions of $n-1$ CPOs $O_2$ and a CPO $O_4$. 
Very recently in \cite{EP} a non-renormalization property of 
the 4-point function of these operators  
was checked to first order in perturbation theory.

For the value of the quartic couplings for the scalars $s^I$
belonging to the massless graviton multiplet we have found \cite{AF6}
the following 2-derivative vertex
\bea
\non
\cL^{(2)}_{AdS5}&=&\frac{2^{14}}{9\pi^3}
C_{I_1I_2I_3I_4}\n_a(s^{I_1}s^{I_2})\n^a(s^{I_3}s^{I_4})
\eea  
and the vertex without derivatives:
\bea
\non
\cL^{(0)}_{AdS5}&=&-\frac{2^{16}}{3\pi^3}
\l C_{I_1I_2I_3I_4}-\frac{1}{6}\d^{I_1I_2}\d^{I_3I_4} \r \times \\
\non
&\times & s^{I_1}s^{I_2}s^{I_3}s^{I_4}.
\eea  
Here the shorthand notation 
$C^{I_1I_2I_3I_4}$ for the trace product of four matrices $C^{I}_{ij}$ was introduced.
Note that $C^{I}_{ij}$ are traceless symmetric matrices whose appearance here 
is due to the explicit description of the spherical harmonics on $S^5$. 
One can also introduce the fields $s_{ij}\equiv C^I_{ij}s^I$ that
provide the natural parametrization of the coset space 
$SL(6,{\bf R})/SO(6)$. Recently we have shown that the relevant part of the gauged
${\cal N}=8$ 5-dimensional supergravity action coincides with the 
action \cite{AF6} for scalars $s^I$ from the massless graviton multiplet. 

\section{Quartic couplings for the extremal case}
In \cite{AF5} we argued that quartic couplings
of the scalars $s^I$ had to vanish in the extremal case when, 
say, $k_1=k_2+k_3+k_4$. This conjecture was based on the fact that 
all exchange Feynman diagrams vanished and 
contact Feynman diagrams had singularity in the extremal case, 
thus non-vanishing quartic couplings would contradict to
the AdS/CFT correspondence. Although the vanishing of the  found quartic 
couplings is not manifest, one can show that 
this important property does take place after an additional field
redefinition. This means that 4-point extremal correlators
of extended CPOs vanish, and also implies the non-renormalization 
theorem \cite{DFMMR} for the corresponding extremal correlators of single-trace CPOs. 
It is clear that since the quartic couplings vanish then 
there should exist such a representation of the quartic couplings, 
that makes the vanishing explicit. An interesting problem is 
to find this representation.

The simplest example of the 4-point 
function of three CPOs $O_2$ and a CPO $O_4$ belongs, actually, to
the class of so-called "sub-extremal" 4-point
functions, for which $k_1=k_2+k_3+k_4-2$. 
The non-renormalization of such correlation functions 
was shown in \cite{EHSSW2} to be the consequence of the superconformal 
Ward identities and of the constrained nature 
of the harmonic superfields.
The non-renormalization theorem also implies the vanishing of 
the corresponding functions of extended CPOs and, since
it is not difficult to show that there is no exchange diagram 
in this case, the corresponding sub-extremal quartic 
couplings of scalars $s^I$ have to vanish too. Just 
recently we have checked that this remarkable 
property indeed takes place. 

\noindent
{\bf Acknowledgements} \hfill\break
The work of G.A. was supported by the the EEC under TMR
contract ERBFMRX-CT96-0045 and in part 
by the Alexander von Humboldt Foundation and by the
RFBI grant N99-01-00166, and the work of S.F. was supported by
the U.S. Department of Energy under grant No. DE-FG02-96ER40967 and
in part by RFBI grant N99-01-00190.

\end{document}